\documentclass[twocolumn,aps,prb,10pt,notitlepage]{revtex4-1}
\usepackage{graphicx,amsmath,color,amssymb,bm,tabularx,comment}
\usepackage[colorlinks=true,linkcolor=cyan,citecolor=cyan]{hyperref}

\newcommand{\n}[0]{\mathbf{x}_\mathbf{n}}
\newcommand{\x}[0]{\mathbf{x}}
\newcommand{\dl}[0]{\mathbf{d}}
\newcommand{\T}[0]{\mathcal{T}}

\newcommand{\Q}[1]{\mathbf{Q}(\mathbf{#1})}
\newcommand{\Qp}[1]{\mathbf{Q}_P(\mathbf{#1})}
\newcommand{\tQ}[1]{\tilde{\mathbf{Q}}(\mathbf{#1})}

\newcommand{\vk}[0]{\mathbf{v}_{m\mathbf{k}}(\n)}

\newcommand{\vkt}[0]{\mathbf{v}_{m\mathbf{k}}(\n,t)}
\newcommand{\cvk}[0]{\tilde{\mathbf{v}}_{m\mathbf{k}}}
\newcommand{\cvkz}[0]{\tilde{\mathbf{v}}_{0\mathbf{k}}}
\newcommand{\cvle}[0]{\tilde{\mathbf{v}}_{m\mathbf{0}}}
\newcommand{\cvlem}[1]{\tilde{\mathbf{v}}_{#1\mathbf{0}}}

\newcommand{\cvkss}[0]{\tilde{\mathbf{v}}_{0\mathbf{0}}}

\newcommand{\wk}[0]{\mathbf{w}_{m\mathbf{k}}(\n)}

\newcommand{\wkr}[0]{\mathbf{w}_{n\mathbf{k'}}(\n)}
\newcommand{\cwk}[0]{\tilde{\mathbf{w}}_{m\mathbf{k}}}
\newcommand{\cwkr}[0]{\tilde{\mathbf{w}}_{n\mathbf{k}}}
\newcommand{\cwkz}[0]{\tilde{\mathbf{w}}_{0\mathbf{k}}}
\newcommand{\cwle}[0]{\tilde{\mathbf{w}}_{m\mathbf{0}}}
\newcommand{\cwlem}[1]{\tilde{\mathbf{w}}_{#1\mathbf{0}}}

\newcommand{\cwkss}[0]{\tilde{\mathbf{w}}_{0\mathbf{0}}}

\newcommand{\lk}[0]{\lambda_{m\mathbf{k}}}

\newcommand{\lkz}[0]{\lambda_{0\mathbf{k}}}
\newcommand{\lkss}[0]{\lambda_{0\mathbf{0}}}

\begin{document}
\title{Reaction-drift-diffusion models from master equations: application to material defects}
\author{Thomas D Swinburne}
\email{swinburne@cinam.univ-mrs.fr}
\affiliation{Universit\'{e} Aix-Marseille, CNRS, CINaM UMR 7325, Campus de Luminy, 13288 Marseille, France}
\author{Danny Perez}
\affiliation{Theoretical Division T-1, Los Alamos National Laboratory, Los Alamos, NM, 87545, USA}
\date{\today}
\begin{abstract}
  We present a general method to produce well-conditioned continuum
  reaction-drift-diffusion equations directly from master equations on a discrete,
  periodic state space. We assume the underlying data to be kinetic Monte Carlo
  models (i.e., continuous-time Markov chains) produced from atomic sampling of point defects in locally periodic environments,
  such as perfect lattices, ordered surface structures or dislocation cores, possibly under
  the influence of a slowly varying external field. Our approach also
  applies to any discrete, periodic Markov chain.
  The analysis identifies a previously omitted non-equilibrium drift term,
  present even in the absence of external forces, which can compete in
  magnitude with the reaction rates, thus being essential
  to correctly capture the kinetics.
  To remove fast modes which hinder time integration, we use a generalized Bloch
  relation to efficiently calculate the eigenspectrum of the master equation.
  A well conditioned continuum equation then emerges by searching for
  spectral gaps in the long wavelength limit, using an established kinetic
  clustering algorithm (e.g., PCCA+) to define a proper reduced state space.
 \end{abstract}
\maketitle
  Multiscale materials modeling is founded on the principle that atomic
  mechanisms at the nanoscale can be `coarse-grained' for use in computationally
  efficient models that reach the mesoscale and
  beyond\cite{kubin2013,dezerald2015first,swinburne2016,alexander2016}.
  As these mechanisms are routinely of outstanding
  complexity\cite{sorensen2000,uberuaga2005,uberuaga2007,perez2009,beland2011},
  they must first be discovered using the best available model of atomic cohesion (either using first
  principle methods such as density functional theory, or with semi-empirical interatomic potentials)
  that computational resources allow. Given a cohesive model, the conceptually simplest approach is to
  generate a molecular dynamics trajectory of the system under study and record transition events
  However, as is well known, the constraint of serial time integration
  means that trajectories generated with molecular dynamics are limited to sub-microsecond timescales,
  often insufficient to observe the rare, thermally activated mechanisms that control the
  drift and diffusion of defects such as vacancy/interstitial clusters, impurity elements
  or adatom islands. This can prevent the accurate up-scaling of atomistic information into accurate
  continuum transport equations.\\

  Over the last three decades a variety of approaches have been developed to overcome
  the timescale issue, including unbiased dynamic
\cite{voter1997,voter1998parallel,TAD,perez2009,chatterjee2015uncertainty,chill2014,perez2016long,swinburne2018b,swinburne2020c}
	or static\cite{dimer,beland2011,wales2002discrete} sampling approaches.
  In the thermally activated regime, any system is extremely likely to thermalize
  in a local energy minima before escaping, giving a well defined separation of
  timescales between vibrations and transitions.
  In this limit, the atomic dynamics can be mapped to a continuous
  time, discrete state Markov chain\cite{Lelievre2018}, which provides the theoretical basis of off-lattice, or atomistic kinetic Monte Carlo (akMC) methods\cite{henkelman2017}, up to an error exponentially small in the
  timescale separation\cite{lebris2012}.
  We have employed this rigorous connection in our massively parallel sampling scheme
  \texttt{TAMMBER}\cite{swinburne2018b,swinburne2020c,tammber}
  (available at \url{github.com/tomswinburne/tammber}), which optimally manages
  many thousands of molecular dynamics `workers' to rapidly discover migration pathways
  of complex defects, with a novel Bayesian metric of sampling completeness
  which can be used to assign well defined uncertainty bounds on the resulting akMC model.
  Further discussion of \texttt{TAMMBER} can be found in a recent review\cite{swinburne2021a}.\\

  The coarse-graining of molecular dynamics to a Markov chain/akMC representation
  is thus well established, with a firm theoretical grounding. However, for all
  but the most simple defects, the corresponding akMC model can be prohibitively cumbersome,
  with a large number of states and huge differences between the fastest and slowest
  transition rates, requiring the use of advanced time-stepping schemes
  \cite{opplestrup2006first,athenes2019elastodiffusion}.
  In addition, whilst the abstract, discrete state space of akMC is sufficiently
  general to allow a formal connection to molecular dynamics, it is clearly
  unsuitable for mesoscale methods such as cluster dynamics
  \cite{jourdan2014efficient,jourdan2015influence,Blondel2017},
  object kinetic Monte Carlo\cite{donev2010first,mason2014,jourdan2021enforcing}
  or phase field models\cite{demange2017prediction,noble2020turing},
  which require a computationally efficient \textit{continuum} representation.
  This is particularly relevant for the growing number of approaches that couple
  point defect dynamics (through a reaction-diffusion equation representation) to
  dislocation dynamics\cite{li2019diffusion,li2020coupled,yu2021coupling}
  or crystal plasticity models\cite{mcelfresh2021discrete}, essential for the modeling of
  irradiation-induced changes in material properties that are central to nuclear materials
  science\cite{swinburne2018c,kohnert2018modeling}.
  As a result, coarse-graining to the mesoscale, either from akMC models or molecular
  dynamics observations, is typically an \textit{ad hoc} phenomenological
  approach, requiring significant end-user expertise to ensure accuracy.\\

  In this paper, we propose a data-driven approach to coarse-grain akMC models into
  reaction-diffusion equations. Following all rigorous coarse-graining schemes
  \cite{pande2010everything}, we exploit separations of timescales in the
  model dynamics, here manifest as a gap in the eigenspectrum of the akMC
  master equation. Compressing the state space to retain only the slow modes then incurs an error exponentially
  small with respect to the size of the gap\cite{lebris2012}. If a sufficiently
  large gap can be found, it is therefore simple to obtain an accurate yet
  dramatically simplified model that allows for efficient time-integration at the continuum scale.\\

  A simple example of a master equation with a spectral gap is that of 'super-basin-to-super-basin`
  dynamics, where a periodically repeating group of states have much faster
  in-group transitions than between-group transitions. This could correspond, e.g., to the evolution
  of a cluster of point defects such that diffusion is slow compared to internal shape fluctuations, in an otherwise periodic crystal.
  A spectral gap then exists which justifies the local equilibrium approximation within each group.
  We show below this reduces the full master equation to a simple Brownian particle
  with an anisotropic diffusivity, with a spatial resolution limited to the lattice constant.
  However, in many settings, the eigenspectrum does not possess a spectral gap,
  implying that a rigorous coarse-graining procedure is not possible. In addition,
  computation of the eigenspectrum can present significant numerical issues.
  We propose a solution, first employing Bloch's theorem to
  efficiently evaluate the reciprocal-space eigenspectrum of periodic
  master equations. We then identify regions of reciprocal space where a spectral
  gap exists, and truncate the short-wavelength modes above the gap, yeilding an
  efficient, coarse-grained model at the expense of a reduced spatial resolution,
  generalizing the simple local equilibrium case. \\

  The paper is structured as follows. We define the master equation for an
  isolated defect under periodic translation symmetries in section \ref{sec:ME}
  and derive the continuum limit. In section \ref{sec:ev} we show how the
  formally infinite dimension eigenvalue problem can be cast in a form
  reminiscient of the Kohn-Sham equations\cite{Martin}, requiring only the
  diagonalization of a small rate matrix over a $\mathbf{k}$-space grid,
  naturally indexing eigenmodes by their spatial frequency and `band'.
  In \ref{sec:sg} the master equation is coarse-grained by leveraging
  timescale separations in the eigenvalue spectrum. As continuum equations
  operate in the limit of slow spatial variation, we need only look for timescale
  separations in this limit, significantly widening the range of systems to which
  such an approach can be applied. The number of below-gap `bands' naturally
  defines the dimension of the reduced state space; the reduced states themselves
  are determined using the PCCA+ method\cite{PCCA,PCCApp}, an established kinetic
  clustering algorithm from the biochemical community\cite{pande2010everything}.
  Our approach, which provides simple reduced models for mesoscale simulations,
  is demonstrated on illustrative toy problems in \ref{sec:app}. Application to
  real systems found in atomistic sampling will be the subject of future work.

\section{Atomic sampling in locally periodic environments}
  In this study, we focus on the kinetics of point defects in a range of
  \textit{locally} periodic geometries, such that the local minima and saddle
  point configurations are invariant under
  translations generated by some primitive set of translations $\T$,
  formed from linear combinations
  of primitive lattice vectors $\mathbf{a}_0,\mathbf{a}_1,\mathbf{a}_2$.
  Examples of translation sets $\T$ for common microstructural environments,
  such as surfaces, near dislocation lines or in the bulk crystal, are shown in table \ref{geom}.
  In this way, we can build exact master equations under the assumption
  of periodic translation symmetry, which yield a continuum
  governing equation with local differential operators.

  This approach can produce local kinetic models for use in more complex microstrutures,
  retaining sufficient complexity to
  exhibit a rich range of behaviors, including
  internal transitions (that can significantly affect
  mobility\cite{uberuaga2005,swinburne2020c}), spontaneous
  disassociation\cite{swinburne2018b}, or interaction with existing
  microstructural features such as dislocations or grain boundaries.

  We note that these symmetries still hold in the presence of
  long-range external stresses that can be considered slowly varying,
  such as elastic interactions\cite{Hudson2005}, a point we return to below.

\section{The master equation}\label{sec:ME}
  To produce a master equation for a translation-symmetric system,
  under no external driving forces,
  we first index each unit cell by a three dimensional discrete lattice position
  $\n=\sum_\alpha\mathrm{n}_\alpha\mathbf{a}_\alpha$, where
  $\mathbf{n}=[\mathrm{n}_0,\mathrm{n}_1,\mathrm{n}_2]$ is an integer vector.
  The occupation probabilities for the $M$ states in a cell at $\n$ are
  contained in a vector $\mathbf{p}(\n)\in\mathbb{R}^{M}$, normalized such that
  $\sum_{\mathbf{n}} \mathbf{1}\cdot\mathbf{p}(\n)=1$,
  where $\mathbf{1}=[1,1,1....]$.\\

  \begin{table}
    \begin{tabularx}{\columnwidth}{l|ll}
      Environment $\quad$ && Translation set $\T$ \\
      \hline
      Bulk crystal && $26$ vectors to adjacent unit cells\\
      Surfaces && $8$ vectors normal to plane \\
      Grain boundaires && $8$ vectors of coincidence site lattice\\
      Dislocations && $2$ vectors $\{\dl,-\dl\}$ along dislocation line
    \end{tabularx}
    \caption{Common environments and corresponding set $\T$ of translation
    vectors for point defects in crystals. Forward and backward directions
    are distinguishable, meaning $\sum_\T\dl=\mathbf{0}$.}
    \label{geom}
  \end{table}

  Transition rates between states in the same cell are contained in a matrix
  $\Q{0}$, with intercell transition rates matrices $\Q{\dl}$ for each $\dl\in\T$.
  With these definitions, the master equation writes
  \begin{equation}
  \frac{\mathrm{d}}{\mathrm{d}t}\mathbf{p}(\n,t)
  =
  [\Q{0}-\mathbf{R}]\mathbf{p}(\n,t)
    + \sum_{\dl\in\T}
     \Q{\dl}\,\mathbf{p}(\n+\dl,t),
     \label{master}
  \end{equation}
  where the diagonal matrix $\mathbf{R}$ contains the total escape rate
  from each state, ensuring probability is conserved, with elements
  \begin{equation}
    \left[\mathbf{R}\right]_{ij}
    =
    \delta_{ij}\left[\mathbf{1}^\top\Q{0}
    +\sum_{\dl\in\T}\mathbf{1}^\top\Q{\dl}\right]_j.
  \end{equation}
  As we use transition state theory, transition rates $k_{ji}$
  satisfy detailed balance\cite{Reichl2009} such that
  $k_{ij}\pi_j=k_{ji}\pi_i$, where $\pi_i$ is the (un-normalized) equilibrium
  occupation probability for state $i$. With a diagonal $M\times M$ matrix
  $[{\bm\Pi}]_{ij} = \delta_{ij}\pi_i$, this implies
  \begin{equation}
    \Q{\dl}{\bm\Pi} = {\bm\Pi}\Q{-\dl}^\top,
    \label{db}
  \end{equation}
  with $\Q{0}{\bm\Pi} = {\bm\Pi}\Q{0}^\top$ being a special case with $\dl=\mathbf{0}$.
  The steady state distribution therefore reads
  \begin{equation}
      \lim_{t\to\infty} \mathbf{p}(\n,t)
      =
      \left(\prod_\alpha{\rm N}_\alpha\right)^{-1}\hat{\bm\pi},
      \label{ss}
  \end{equation}
  where ${\rm N}_\alpha$ is the total number of cells in each primitive direction
  and $\hat{\bm\pi}$ is the cell normalized Boltzmann distribution.

\subsubsection{External driving force}
    In the presence of a constant external driving force $\mathbf{f}$,
    typically due to a slowly varying stress gradient or an external electric field for
    charged defects,
    the detailed balance equation is modified to become
    \begin{equation}
        \Q{\dl,{\bf f}}{\bm\Pi}
        \exp\left({\beta\dl\cdot{\bf f}}\right)
        = {\bm\Pi}\Q{-\dl,{\bf f}}^\top,
        \label{gdb}
    \end{equation}
    where $\beta=1/{\rm k_BT}$, $\Q{\dl,{\bf f}}$ are modified rate matrices
    and $\dl\cdot{\bf f}$ is the additional energy difference
    between initial and final states due to ${\bf f}$.
  The steady state distribution becomes
  \begin{equation}
      \lim_{t\to\infty} \mathbf{p}(\n,t)
      =
      \left(\prod_\alpha
      \frac
      {1-e^{-{\beta\mathbf{a}_\alpha\cdot{\bf f}}}}
      {{1-e^{-\beta{\rm N}_\alpha \mathbf{a}_\alpha\cdot{\bf f}}}}
      \right)
      e^{-\beta\n\cdot{\bf f}}
    \hat{\bm\pi},
  \end{equation}
  which reduces to (\ref{ss}) as $|\mathbf{f}|\to0$.
  Under the widely used midpoint rule\cite{}, the modified rate matrices are given by
  \begin{equation}
        \Q{\dl,{\bf f}} \simeq \exp\left(-{\beta\dl\cdot{\bf f}}/2\right)\Q{\dl},
  \end{equation}
  which clearly satisfies (\ref{gdb}).


\subsection{Continuum Limit}\label{sec:cl}
  The continuum limit of (\ref{master}) is defined under the assumption that
  occupation probabilities vary slowly on the length scale of a single unit cell.
  In the next section, where we use Bloch's theorem to diagonalise (\ref{master}),
  this is equivalent to only considering long-wavelength modes.
  In this limit, we can equate the cell probability vector $\mathbf{p}(\n,t)$ to a
  continuous (vector) probability density ${\bm\rho}(\x,t)$, which varies slowly over
  a unit cell of volume
  $\mathrm{V}=\mathbf{a}_0\cdot(\mathbf{a}_1\times\mathbf{a}_2)$, giving
  \begin{equation}
    \mathbf{p}(\n,t)\to {\bm\rho}(\x,t)\mathrm{V}\mathrm{d}^3\x.
    \label{continuum}
  \end{equation}
  Under the same assumption of slow variation, we make a Taylor expansion of
  ${\bm\rho}(\x,t)$ to second order. The master equation (\ref{master}) becomes
  the reaction-diffusion equation
  \begin{equation}
    \frac{\mathrm{d}}{\mathrm{d}t}{\bm\rho}(\x,t)
    =
    \mathbb{Q}{\bm\rho}(\x,t)
    +
    \mathbb{C}{\bm\rho}(\x,t)
    +
    \mathbb{D}{\bm\rho}(\x,t),
    \label{rde}
  \end{equation}
  where, the reaction, drift and diffusion terms read
  \begin{align}
    \mathbb{Q} &= \Q{0} + \sum_{\dl\in\T}\Q{\dl} - \mathbf{R}
    ,\\
    \mathbb{C} &= \sum_{\dl\in\T}\Q{\dl} \left(\dl\cdot{\bm\nabla}\right)
    ,\\
    \mathbb{D} &= \frac{1}{2} \sum_{\dl\in\T}\Q{\dl}
    \left(\dl\cdot{\bm\nabla}\right)^2.
    \label{operators}
  \end{align}
  The master equation (\ref{master}) and its continuum limit (\ref{rde})
  are the central equations of this paper.

  We note that conventional reaction-diffusion equations
  assume the drift term $\mathbb{C}{\bm\rho}$ to vanish; however,
  this is only true when a local cell-wise equilibrium approximation
  ${\bm\rho}(\x,t)\simeq\hat{\bm\pi}\rho(\x,t)$ is valid and
  detailed balance holds, as we show below.In the general case, $\mathbb{C}{\bm\rho}$ can play an important role
  in non-equilibrium mixing physics. In the examples below we demonstrate
  that these drift terms can be of greater magnitude than the pure reaction
  rates.


\section{Simplifying the master equation}\label{sec:ev}
  As discussed in the introduction, whilst the master equation (\ref{master})
  is exact under suitable limits, the number of states $M$ can be very large,
  and the rate matrices can have widely varying rates, potentially introducing severe stability
  constraints on the integration timestep. This complexity
  will similarly affect the continuum limit (\ref{rde}).

  In this section, we first use Bloch's theorem\cite{Ashcroft} to derive a closed form for the
  eigenmodes of the master equation (\ref{master}), requiring only the diagonalization of small
  $M\times M$ matrices, returning eigenvectors and eigenvalues which can naturally be indexed
  with respect to a reciprocal space vector $\mathbf{k}$ in the first Brillouin zone and a
  `band' index $m\in[1,M]$. Access to all the eigenvalues then allows us to look for gaps in the eigenvalue spectrum,
  with all above-gap modes considered `fast' as they decay on a timescale where `slow'
  modes are essentially constant.   By removing the `fast' modes we produce an effective master equation with a smaller number
  of states in each cell, equal to the number of retained bands, with a much narrower range of
  intrinsic timescales.

  In cases where we find only one state per cell, this procedure results in a simple anisotropic diffusion equation.
  In the general case, where a spectral gap may not be observed, we provide a solution
  for the continuum limit (\ref{rde}), which is equivalent to only considering the dynamics
  of long wavelength modes $|\mathbf{k}|\to0$. We show that the long-wavelength spectrum
  often reveals effective spectral gaps at the cost of reduced spatial resolution, which
  we use to produce well-conditioned continuum equations.

\subsection{Eigenmodes of the master equation}
  As each cell probability vector $\mathbf{p}(\n)$ is coupled to neighboring cells
  in the master equation (\ref{master}), the effective transition rate matrix
  is of infinite dimension for an infinite system. However, due to the periodic translation symmetry,
  a solvable form of (\ref{master}) can be obtained using Bloch's theorem\cite{Ashcroft}.
  We first define lattice translation operators
  \begin{equation}
    \hat{\mathbf{T}}(\mathbf{x}_{\mathbf{m}}) \mathbf{p}(\n)
    \equiv
    \mathbf{p}(\n+\mathbf{x}_{\mathbf{m}}), \label{trans_op}
  \end{equation}
  to write the master equation (\ref{master}) in the operator form
  \begin{equation}
    \frac{\mathrm{d}}{\mathrm{d}t}\mathbf{p}(\n)
    =
    \hat{\mathbf{Q}}\mathbf{p}(\n)
    ,\quad
    \hat{\mathbf{Q}}
    \equiv
    \Q{0}-\mathbf{R}+\sum_{\dl\in\T}\Q{\dl}
    \hat{\mathbf{T}}(\dl).\nonumber
  \end{equation}
  Our goal is to find the (negated) eigenvalues and eigenfunctions
  \begin{equation}
    \hat{\mathbf{Q}}\mathbf{v}(\n) = -\lambda\mathbf{v}(\n)
    ,\quad
    \mathbf{v}(\n,t) = \exp(-\lambda t)\mathbf{v}(\n),
    \label{gen_ev}
  \end{equation}
  where definition of the sign of $\lambda$ reflects the fact that
  $\lambda\geq0$ for physical Master equations.

  As $\hat{\mathbf{Q}}$ clearly commutes with any
  combination of $\hat{\mathbf{T}}(\dl)$, its vector-valued
  eigenfunctions satisfy a Bloch relation\cite{Ashcroft}
  \begin{equation}
    \vk \equiv
    \exp(\mathrm{i}\n\cdot\mathbf{k})
    \cvk,
    \label{evbloch}
  \end{equation}
  where $m\in[1,M]$, $\mathbf{k}$ is a wavevector in the first Brillouin
  zone of the lattice defined by $\T$, and $\cvk\in\mathbb{R}^M$ is constant
  across cells (but not within).
  The eigenvalue problem reduces to
  \begin{equation}
    \tQ{k}\cvk
    =
    -\lk
    \cvk,
    \label{bloch_evp}
  \end{equation}
  where the $M\times{M}$ `Bloch' transition matrix $\tQ{k}$ writes
  \begin{equation}
    \tQ{k}
    \equiv
    \Q{0}-\mathbf{R}+\sum_{\dl\in\T}\Q{\dl}
    \exp(\mathrm{i}\dl\cdot\mathbf{k}).
    \label{blochM}
  \end{equation}
  Computation of eigenvalues and eigenmodes therefore requires diagonalizing
  an $M\times M$ matrix $\tQ{k}$ across some discrete grid
  of $\mathbf{k}$-points in the first Brillouin zone, in close analogy to
  Kohn-Sham density functional theory\cite{Martin}.

  This procedure will return right eigenvectors $\cvk$ and eigenvalues $\lk$
  as defined above, in addition to left eigenvectors $\cwk$,
  which satisfy the generalized orthogonality condition
  $\cwkr\cdot\cvk=\delta_{nm}$. The master equation has
  left Bloch eigenvectors $\wk = \exp(-\mathrm{i}\n\cdot\mathbf{k})\cwk$,
  such that $\sum_{\n}\wkr\cdot\vk = N_V\delta_{nm}\delta_{\mathbf{k'k}}$,
  where $N_V$ is the total number of cells in the system.

  The general time evolution of probability then reads
  \begin{equation}
    \mathbf{p}(\n,t)
    =
    \sum_{\mathbf{k},m}
    a_{m\mathbf{k}}
    \vkt
    ,\label{expansion}
  \end{equation}
  where $\vkt=\vk\exp(-\lk t)$ and $a_{m\mathbf{k}}=\sum_{\n}\wk\cdot\mathbf{p}(\n,0)$.
  As probability distributions
  are real, it is simple to show that $a^\dag_{m,-\mathbf{k}} = a_{m\mathbf{k}}$.
  The steady state has eigenvalue $\lkss=0$, with cell eigenvectors
  \cite{swinburne2020d}
  \begin{equation}
    \cwkss = \mathbf{1}
    ,\quad
    \cvkss = \hat{\bm\pi}
    ,\quad
    a_{0\mathbf{0}}=\left(\prod_\alpha{\rm N}_\alpha\right)^{-1},
  \end{equation}
  in agreement with (\ref{ss}).

  \subsubsection{External driving force}
  The Bloch relation (\ref{evbloch}) derives from the general form of the
  eigenfunctions of the $\hat{\mathbf{T}}(\dl)$. As these periodic translation
  operators form a representation of the cyclic group\cite{scott2012group},
  it can be shown that the eigenvalues must be $M$-fold degenerate and
  be a scalar representation of the cyclic group, namely phase factors
  $\exp(\mathrm{i}\dl\cdot\mathbf{k})$.

  In the presence of an external driving force ${\bf f}$, the
  modified detailed balance relations
  (\ref{gdb}) imply that $\hat{\mathbf{Q}}$ now commutes with
  $\exp(-\beta\dl\cdot{\bf f})\hat{\mathbf{T}}(\dl)$. As the
  scalars $\phi(\dl)=\exp(-\beta\dl\cdot{\bf f})$ have exactly the same
  commutation relations as the $\hat{\mathbf{T}}(\dl)$,
  the eigenfunctions of this modified operator are of exactly the same form,
  except that the phase factor also accounts for the energy difference due to
  $\bf f$, giving a generalized Bloch relation
  \begin{equation}
    \vk \equiv
    \exp(\n\cdot[\mathrm{i}\mathbf{k}-\beta\mathbf{f}])
    \cvk.
  \end{equation}
  The full eigenspectrum can be found by solving the eigenvalue problem (\ref{bloch_evp})
  as above, with suitably modified rate matrices satisfying (\ref{gdb}).

\subsection{Spectral coarse-graining}\label{sec:sg}
  The general principle behind spectral coarse-graining is to look for a spectral gap in the ordered list
  of (negated) eigenvalues $\{\lk\}$ . Aside from the steady state $\lambda_{0\mathbf{0}}=0$,
  all modes are exponentially decaying with $\lk>0$.
  For any timescale $\tau$, we are free to define a (possibly vanishing)
  spectral gap $\Delta_\tau>0$ through
  \begin{equation}
    \Delta_\tau \equiv
    \left(\min_{\lk\tau\geq1}\lk\right)
    -
    \left(\max_{\lk\tau<1}\lk\right).
  \end{equation}
  After a time $t\sim\tau$ the general expansion
  (\ref{expansion}) reads
  \begin{align}
    \mathbf{p}(\n,t)
    &\to
    \sum_{\lk\tau<1}
    a_{m\mathbf{k}}
    \vkt
    +
    \mathcal{O}(\exp(-\tau\Delta_\tau)).
    \nonumber
  \end{align}
  If $\tau\Delta_\tau\gg1$, the reduced order model formed from below-gap modes
  will therefore have exponentially small error with respect to the true dynamics. If the
  number of below-gap modes is small, the reduced order model is typically
  much better suited for continuum implementation. Our objective is therefore
  to find conditions under which such a gap exists.

\subsection{Continuum limit in reciprocal space}
  The continuum limit (\ref{rde}) is defined to apply only to probability
  densities which are initialized to be slowly varying on the length-scale of a unit cell.
  From the Bloch relation (\ref{evbloch}), it is clear that this is
  equivalent to only considering contributions from eigenmodes of
  long spatial wavelength, where $|\mathbf{k}\cdot\dl| \ll 1$.
  To produce a well-conditioned continuum equation, we
  thus do not need to have a spectral gap in the entire
  eigenmode distribution. We can instead focus on the $M$ bands $\lk,m\in[1,M]$,
  and look for the long wavelength spectral gap
  \begin{equation}
      \lambda_{(P+1)\mathbf{k}}\gg\lambda_{P\mathbf{k}}
      ,\quad
      |\mathbf{k}\cdot\dl| \ll 1.
  \end{equation}
  To determine the spatial resolution more precisely, we can use
  first order perturbation theory to evaluate the long wavelength
  eigenvalue approximation
  \begin{equation}
      \lk
      \simeq
      \lk^0 =
      -\cwle^\top\tQ{k}\cvle
      ,\quad
      |\mathbf{k}\cdot\dl| \ll 1,
      \label{lwv_ev}
  \end{equation}
  where $\{\cwle,\cvle\}$ form the $\mathbf{k}=\mathbf{0}$ basis.
  An appropriate grid size $L$ can then
  be determined through the error measure
  \begin{equation}
    |\lambda_{m\mathbf{k}^L}^0-\lambda_{m\mathbf{k}^L}|
    =\epsilon|\lambda_{m\mathbf{k}^L}|
    ,\quad|\mathbf{k}^L\cdot\mathbf{1}|_{\infty}=\frac{\pi}{L},
  \end{equation}
  where $|{\bf x}|_{\infty} \equiv \max(|{\rm x}_0|,|{\rm x}_1|,\dots)$.
  If such a gap exists, we can produce an accurate reaction-diffusion
  equation from a reduced number states per cell, equal to the number of retained
  bands, $P<M$, with a corresponding compression in the range of inherent timescales, which makes time integration
  significantly more efficient.

\subsection{Building coarse-grained states with PCCA+}
  To determine the coarse grained states, we wish to partition the $M$ states into
  $P$ groups such that the intergroup transition rates are minimized, to ensure the
  timescale compression. This is a common objective in Markov model analysis;
  a popular approach, used here, is the Robust Perron Cluster Analysis (PCCA+)
  technique\cite{PCCApp}, which is widely used in the biochemical community\cite{pande2010everything}.
  Whilst the original PCCA method\cite{PCCA} attempts
  to find a strict (non-overlapping) partitioning of the $M$ states into $P$
  coarse-grained states, the PCCA+ technique instead forms `fuzzy' clusters from
  linear combinations of the $P$ slow eigenmodes. This is ideal for our usage, as
  we can then construct coarse-grained transition state matrices for use in the
  continuum limit (\ref{rde}). We employ a python implementation of the PCCA+
  algorithm from the \texttt{MSMTools} package\cite{msmtools}.


  To define our coarse-grained states, we first form matrices of
  the slow eigenvectors
  \begin{align}
    \mathbf{W}_P
    &\equiv
    [\cwlem{0},\dots,\cwlem{P-1,}]\in\mathbb{R}^{P\times M}
    ,\nonumber\\
    \mathbf{V}_P&\equiv[\cvlem{0},\dots,\cvlem{P-1,}]\in\mathbb{R}^{P\times M},
  \end{align}
  The PCCA+ algorithm is then used to build the $P\times P$ clustering matrix
  of linear coefficients
  \begin{equation}
    \mathbf{M}_P = \texttt{PCCA+}(\Q{0}) \in \mathbb{R}^{P\times P},
  \end{equation}
  which gives a coarse-grained probability vector
  \begin{equation}
    {\bm\rho}_P(\x,t) \equiv \mathbf{M}_P\mathbf{W}_P{\bm\rho} \in\mathbb{R}^P,
  \end{equation}
  with coarse grained transition matrices
  \begin{equation}
    \Qp{d} \equiv \mathbf{M}_P\mathbf{W}_P\Q{d}\mathbf{V}^\top_P\mathbf{M}^\top_P,
  \end{equation}
  from which we can build a coarse grained reaction-diffusion equation
  operators $\mathbb{Q}_P,\mathbb{C}_P,\mathbb{D}_P$,
  in direct analogy with (\ref{rde})
  \begin{equation}
     \frac{\partial}{\partial t}{\bm\rho}_P(\x,t)
     =
     \mathbb{Q}_P{\bm\rho}_P+
     \mathbb{C}_P{\bm\rho}_P
     +\mathbb{D}_P{\bm\rho}_P.
  \end{equation}
  In the next section, we apply this methodology to illustrative examples.

  \section{Example applications}\label{sec:app}

  \subsection{Simplest case: local equilibrium, P=1}
  We first treat the simplest case, where the lowest eigenvalue band
  is well separated from all the others, corresponding to the case where
  intracell transitions are much faster than intercell transitions.
  The left and right eigenvectors $\cwkz,\cvkz$ will be to a high degree
  of approximation the steady state vectors $\mathbf{1},\hat{\bm\pi}$
  for all values of $\mathbf{k}$, meaning the slow eigenmodes can be written
   \begin{equation}
    \cwkz \simeq \mathbf{1},\,
    \cvkz \simeq \hat{\bm\pi},
    \nonumber
  \end{equation}
  where these relations become exact in the limit of vanishing intercell transitions rates.
  As we have only one cluster, our reduced probability density is the scalar
  \begin{equation}
       {\rho}_P(\x,t)= \mathbf{1}\cdot\rho(\x,t)
       ,
  \end{equation}
  which is clearly a local equilibrium approximation.
  This immediately implies that the reaction term vanishes-
  \begin{equation}
      \mathbb{Q}_P{\rho}_P(\x,t) = \mathbf{1}^\top\left[\sum_{\dl\in\T}\Q{d}-\mathbf{R}\right]\hat{\bm\pi}\rho(\x,t) = 0.
  \end{equation}
  The long wavelength approximation (\ref{lwv_ev}) is then highly accurate,
  with eigenvalues of the form
  \begin{equation}
    \lkz \simeq \lkz^0
    =
    \sum_{\dl\in\T}
    2\sin^2\left(\dl\cdot\mathbf{k}\right)
    \left(\mathbf{1}^\top\Q{d}\hat{\bm\pi}\right).
  \end{equation}

  As mentioned above, the drift term vanishes due to the detailed balance symmetry (\ref{db})
  \begin{equation}
    \mathbb{C}_P{\rho}_P(\x,t) =
    \frac{1}{2}
    \sum_{\dl\in\T}
    \mathbf{1}^\top\left(
    \Q{d}
    -\Q{-d}
    \right)\hat{\bm\pi}\dl\cdot{\bm\nabla}{\rho}_P
    ={0},\nonumber
  \end{equation}
  whilst the diffusion operator acts only on the scalar probability density
  $\rho(\x,t)$
  \begin{equation}
    \mathbb{D}_P{\rho}_P(\x,t) = {\bm\nabla}\cdot\mathbf{D}\cdot{\bm\nabla}\rho_P(\x,t).
  \end{equation}
  We can therefore take the continuum limit (\ref{rde}) for $\rho$,
  which has the pure diffusion form
  \begin{equation}
  \frac{\mathrm{d}}{\mathrm{d}t}\rho(\x,t)
  =
  {\bm\nabla}\cdot\mathbf{D}\cdot{\bm\nabla}\rho(\x,t)
  \end{equation}
  where the $3\times3$ diffusion matrix reads, from (\ref{rde}),
  \begin{equation}
    \mathbf{D} = \frac{1}{2}\sum_{\dl\in\T}
    \left(\mathbf{1}^\top\Q{d}\hat{\bm\pi}\right)
    \dl\dl^\top.
    \label{Deff}
  \end{equation}
  For sufficiently symmetric defects the diffusion matrix
  can reduce to $\mathbf{D}=\mathrm{D}\mathbb{I}$, giving
  the familiar isotropic diffusion equation
  $({\mathrm{d}}/{\mathrm{d}t}){\rho} = \mathrm{D}\nabla^2\rho$.
  An example of such a system is shown in figure \ref{toy}a, where we
  have labeled the transition matrices $\Q{d}$.

  \subsection{Long wavelength spectral gap: P=1}
\begin{figure}[!th]
  \centering
  \includegraphics[width=\columnwidth]{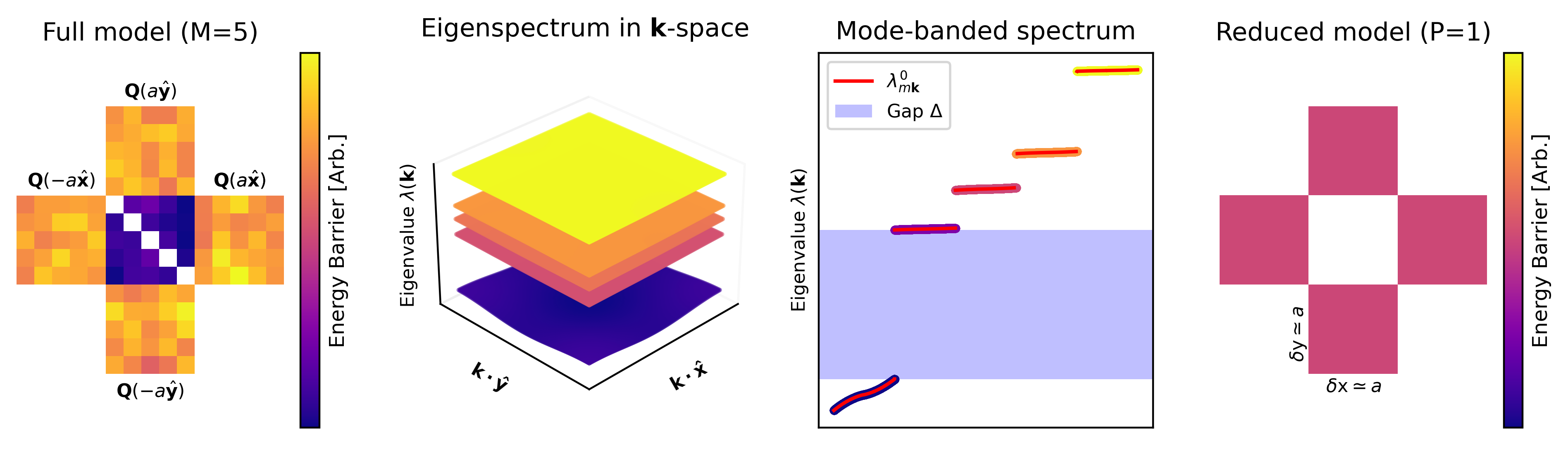}
  \includegraphics[width=\columnwidth]{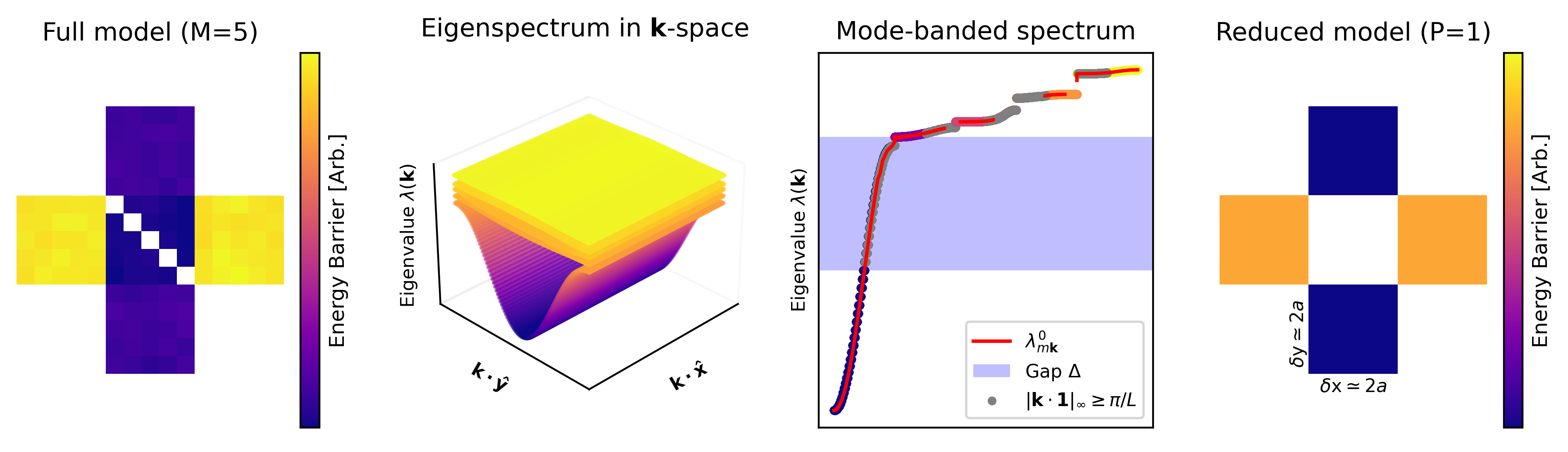}
  \includegraphics[width=\columnwidth]{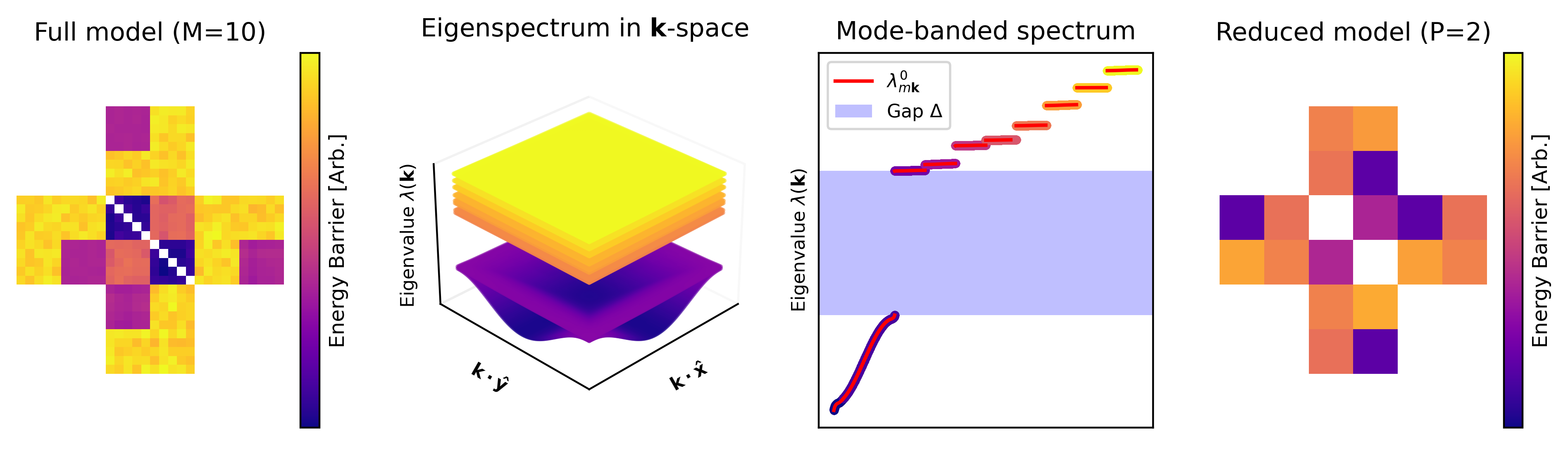}
  \caption{Coarse graining of a model systems which
  diffuse in two dimensions. For simplicity we only consider
  nearest neighbor transitions. Top: A simple `mixing' case, where local
  equilibrium is reached in the cell before any transitions.
  The full eigenspectrum has a clear gap; retaining only the
  slow modes returns a simple single state diffusion model.
  Middle: A more complex case, where the timescale for diffusion along
  $\mathbf{y}$ is comparable to that of intercell
  mixing. Whilst the full spectrum is dense, we can find a gap in a
  restricted region of $\mathbf{k}$-space, corresponding to a coarser spatial
  resolution in the fast diffusion direction, returning a reduced order model.
  Bottom: A multi-state case, where the cell divides into two superbasins, each
  having fast migration kinetics along different directions.
  A spectral gap exists above the second band,
  returning a two state model with the PCCA+ procedure. }
  \label{toy}
\end{figure}

  As a slightly more complex case, consider the system in figure \ref{toy}b,
  which has a clear separation of timescales for migration in one direction
  but competing timescales for mixing and migration in the other.
  However, for the continuum equation, we can `create' a spectral gap by
  setting the spatial resolution through an upper bound on $\mathbf{k}$,
  such that the largest permitted mode of the highest retained band still has
  a well defined gap to the lowest mode of the next band. A larger gap,
  and thus a lower error in the propagated density, can be found by
  sacrificing spatial resolution, offering a simple trade-off depending on
  the desired application. In the case presented in figure \ref{toy}b we
  recover the same simple diffusion form as presented in (\ref{Deff}),
  but with a highly anisotropic diffusivity.

  \subsection{Multistate reaction-diffusion : P=2}
  We now consider additional structure in the model, designed to return a two-state
  reaction-diffusion equation. The inter-cell transitions split into two super-basins,
  with one super-basin having fast migration along $\hat{\mathbf{x}}$, and the other
  having fast migration along $\hat{\mathbf{y}}$. The long-time behaviour is thus
  isotropic, but the short-time behaviour is anisotropic.
  A physical example of such a system is the crowdion
  self-interstitial defect found in some body centered cubic metals
  \cite{swinburne2014,swinburne2017}, which executes
  long periods of fast one dimensional migration along a given
  $\langle111\rangle$ direction, punctuated by rare rotations between different symmetry-equivalent directions.

  With $P$ (in this example $P=2$) retained bands, we construct an effective
  coarse-grained system of $P$ effective states using the PCCA+ technique described above.
  However, in the present simplified example the PCCA+ clustering
  can be performed analytically in the limit of strongly metastable superbasins,
  where the second-lowest left eigenvector $\cwlem{1}$ will
  be approximately $-1$ in one basin and $1$ in the other. As $\cwlem{0}=\mathbf{1}$ is
  constant, we can then select one basin with $(\cwlem{0}+\cwlem{1})/2$ and the other
  with $(\cwlem{0}-\cwlem{1})/2$, giving an approximate clustering matrix
  \begin{equation}
     \mathbf{M}_P \simeq  \frac{1}{2}\left[\begin{matrix}
                        1 & 1 \\
                        -1 & 1
                    \end{matrix}\right].
  \end{equation}
  This form is can be expected for strongly metastable dynamics when $P=2$.

  To investigate the form of the resultant continuum equation, we
  expand the reduced reaction, drift and diffusion operators into
  $2\times2$ matrix-valued coefficients
  \begin{align}
      \mathbb{Q}_P{\bm\rho}_P
      &=
      \mathbf{Q}_P{\bm\rho}_P
      ,\quad
      \mathbb{C}_P{\bm\rho}_P
      =
      \sum_\alpha
      \mathbf{C}_\alpha\frac{\partial}{\partial{\rm x}_\alpha}{\bm\rho}_P \\
      \mathbb{D}_P{\bm\rho}_P
      &=
      \sum_{\alpha,\alpha'}
      \mathbf{D}_{\alpha\alpha'}\frac{\partial^2}{\partial{\rm x}_\alpha\partial{\rm x}_{\alpha'}}{\bm\rho}_P.
  \end{align}
 by carrying out the sums over the primitive set of translations $\T$ in the coarse-grained versions of Eq.\ \ref{operators} and gathering the terms corresponding to the same derivatives.
  From the form of the model presented in figure (\ref{toy}), one can expect that
  the diffusion coefficients $\mathbf{D}_{\alpha\alpha'}$  will have the largest
  components, due to the fast migration directions along either $\mathbf{x}$ or $\mathbf{y}$.

\begin{figure}
    \centering
    \includegraphics[width=\columnwidth]{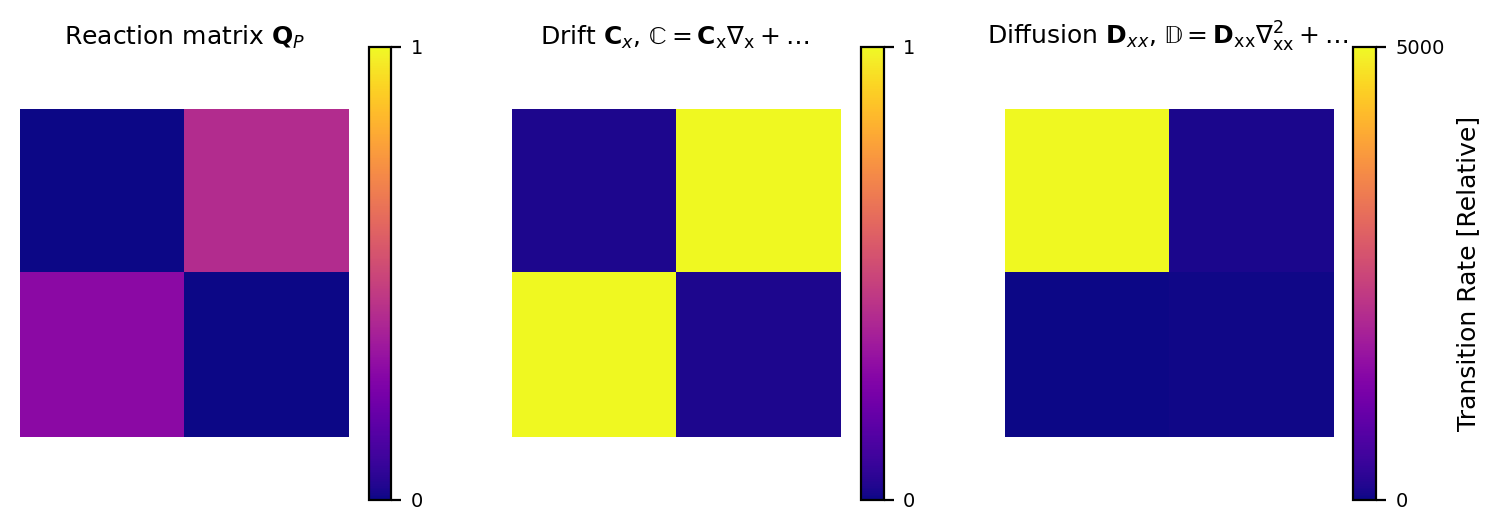}
    \caption{Some matrix coefficients forming operators in the reaction-diffusion equation
    for the last example shown in figure \ref{toy}. Matrix elements are colored on a relative scale
    to aid comparison. The drift coefficients are comparable in magnitude to the
    reaction terms, showing that the typically neglected drift operator $\mathbb{C}$ is
    essential to capture the correct kinetics.}
    \label{toy2}
\end{figure}

  However, to change migration direction, the system must execute a inter-basin transition
  (i.e. between coarse grained states), either in the same cell, as governed by the off-diagonal components of $\mathbf{Q}_P$, $\mathbf{C}_\alpha$ and $\mathbf{D}_{\alpha\alpha'}$.
  In figure \ref{toy2} we plot these coefficient matrices for the reaction operator and
  drift or diffusion along $\mathbf{x}$. As can be seen, the diffusion matrix has essentially
  one non-zero component, on the diagonal, corresponding to transitions between periodically
  equivalent coarse-grained states. Notably, the drift operator, missing in standard
  derivations of reaction-diffusion equations, is of greater magnitude than the
  coefficients in the reaction matrix, showing that these terms must not be neglected in general
  when building multi-state reaction-diffusion equations.

\section{Conclusions}
In this paper, we have developed a comprehensive and efficient approach to
coarse-grain complex periodic continuum Markov chains into compact, accurate, and numerically efficient
continuum reaction-drift-diffusion equations. To do so, we
calculate the eigenmodes of general master equations for materials defects
employing a generalized Bloch theorem. Access to this spectrum is then used with
an established kinetic clustering routine, the PCCA+ method\cite{PCCApp},
to build well-conditioned continuum reaction-drift-diffusion equations.
In future work this methodology will be applied to data from atomic sampling
and implemented in mesoscale modeling schemes.
\section{Code Availability}
Code to generate the figures can be found at \url{github.com/tomswinburne/ReactionDiffusion}.
\section{Acknowledgements}
TDS gratefully recognizes support from the Agence Nationale de Recherche, via the MEMOPAS project ANR-19-CE46-0006-1. This work was granted access to the HPC resources of IDRIS under the allocation A0090910965 attributed by GENCI, and has been carried out within the framework of the EUROfusion consortium and has received funding from the Euratom research and training programme 2019-2020 under grant agreement No 633053. The views and opinions expressed herein do not necessarily reflect those of the European Commission.
DP was supported by the Laboratory Directed Research and Development program of Los Alamos National Laboratory under project number 20220063DR.
Los Alamos National Laboratory is operated by Triad National Security LLC, for the National Nuclear Security administration of the U.S. DOE under Contract No. 89233218CNA0000001.

\bibliography{used.bib}

\begin{thebibliography}{51}%
\makeatletter
\providecommand \@ifxundefined [1]{%
 \@ifx{#1\undefined}
}%
\providecommand \@ifnum [1]{%
 \ifnum #1\expandafter \@firstoftwo
 \else \expandafter \@secondoftwo
 \fi
}%
\providecommand \@ifx [1]{%
 \ifx #1\expandafter \@firstoftwo
 \else \expandafter \@secondoftwo
 \fi
}%
\providecommand \natexlab [1]{#1}%
\providecommand \enquote  [1]{``#1''}%
\providecommand \bibnamefont  [1]{#1}%
\providecommand \bibfnamefont [1]{#1}%
\providecommand \citenamefont [1]{#1}%
\providecommand \href@noop [0]{\@secondoftwo}%
\providecommand \href [0]{\begingroup \@sanitize@url \@href}%
\providecommand \@href[1]{\@@startlink{#1}\@@href}%
\providecommand \@@href[1]{\endgroup#1\@@endlink}%
\providecommand \@sanitize@url [0]{\catcode `\\12\catcode `\$12\catcode
  `\&12\catcode `\#12\catcode `\^12\catcode `\_12\catcode `\%12\relax}%
\providecommand \@@startlink[1]{}%
\providecommand \@@endlink[0]{}%
\providecommand \url  [0]{\begingroup\@sanitize@url \@url }%
\providecommand \@url [1]{\endgroup\@href {#1}{\urlprefix }}%
\providecommand \urlprefix  [0]{URL }%
\providecommand \Eprint [0]{\href }%
\providecommand \doibase [0]{http://dx.doi.org/}%
\providecommand \selectlanguage [0]{\@gobble}%
\providecommand \bibinfo  [0]{\@secondoftwo}%
\providecommand \bibfield  [0]{\@secondoftwo}%
\providecommand \translation [1]{[#1]}%
\providecommand \BibitemOpen [0]{}%
\providecommand \bibitemStop [0]{}%
\providecommand \bibitemNoStop [0]{.\EOS\space}%
\providecommand \EOS [0]{\spacefactor3000\relax}%
\providecommand \BibitemShut  [1]{\csname bibitem#1\endcsname}%
\let\auto@bib@innerbib\@empty
\bibitem [{\citenamefont {Kubin}(2013)}]{kubin2013}%
  \BibitemOpen
  \bibfield  {author} {\bibinfo {author} {\bibfnamefont {L.}~\bibnamefont
  {Kubin}},\ }\href {http://books.google.co.uk/books?id=Uit4KrwBmrwC} {\emph
  {\bibinfo {title} {{Dislocations, Mesoscale Simulations and Plastic
  Flow}}}},\ Oxford Series on Materials Modelling\ (\bibinfo  {publisher} {OUP
  Oxford},\ \bibinfo {year} {2013})\BibitemShut {NoStop}%
\bibitem [{\citenamefont {Dezerald}\ \emph {et~al.}(2015)\citenamefont
  {Dezerald}, \citenamefont {Proville}, \citenamefont {Ventelon}, \citenamefont
  {Willaime},\ and\ \citenamefont {Rodney}}]{dezerald2015first}%
  \BibitemOpen
  \bibfield  {author} {\bibinfo {author} {\bibfnamefont {L.}~\bibnamefont
  {Dezerald}}, \bibinfo {author} {\bibfnamefont {L.}~\bibnamefont {Proville}},
  \bibinfo {author} {\bibfnamefont {L.}~\bibnamefont {Ventelon}}, \bibinfo
  {author} {\bibfnamefont {F.}~\bibnamefont {Willaime}}, \ and\ \bibinfo
  {author} {\bibfnamefont {D.}~\bibnamefont {Rodney}},\ }\href@noop {}
  {\bibfield  {journal} {\bibinfo  {journal} {Physical Review B}\ }\textbf
  {\bibinfo {volume} {91}},\ \bibinfo {pages} {094105} (\bibinfo {year}
  {2015})}\BibitemShut {NoStop}%
\bibitem [{\citenamefont {Swinburne}\ \emph {et~al.}(2016)\citenamefont
  {Swinburne}, \citenamefont {Arakawa}, \citenamefont {Mori}, \citenamefont
  {Yasuda}, \citenamefont {Isshiki}, \citenamefont {Mimura}, \citenamefont
  {Uchikoshi},\ and\ \citenamefont {Dudarev}}]{swinburne2016}%
  \BibitemOpen
  \bibfield  {author} {\bibinfo {author} {\bibfnamefont {T.~D.}\ \bibnamefont
  {Swinburne}}, \bibinfo {author} {\bibfnamefont {K.}~\bibnamefont {Arakawa}},
  \bibinfo {author} {\bibfnamefont {H.}~\bibnamefont {Mori}}, \bibinfo {author}
  {\bibfnamefont {H.}~\bibnamefont {Yasuda}}, \bibinfo {author} {\bibfnamefont
  {M.}~\bibnamefont {Isshiki}}, \bibinfo {author} {\bibfnamefont
  {K.}~\bibnamefont {Mimura}}, \bibinfo {author} {\bibfnamefont
  {M.}~\bibnamefont {Uchikoshi}}, \ and\ \bibinfo {author} {\bibfnamefont
  {S.~L.}\ \bibnamefont {Dudarev}},\ }\href@noop {} {\bibfield  {journal}
  {\bibinfo  {journal} {Scientific Reports}\ }\textbf {\bibinfo {volume} {6}}
  (\bibinfo {year} {2016})}\BibitemShut {NoStop}%
\bibitem [{\citenamefont {Alexander}\ \emph {et~al.}(2016)\citenamefont
  {Alexander}, \citenamefont {Marinica}, \citenamefont {Proville},
  \citenamefont {Willaime}, \citenamefont {Arakawa}, \citenamefont {Gilbert},\
  and\ \citenamefont {Dudarev}}]{alexander2016}%
  \BibitemOpen
  \bibfield  {author} {\bibinfo {author} {\bibfnamefont {R.}~\bibnamefont
  {Alexander}}, \bibinfo {author} {\bibfnamefont {M.-C.}\ \bibnamefont
  {Marinica}}, \bibinfo {author} {\bibfnamefont {L.}~\bibnamefont {Proville}},
  \bibinfo {author} {\bibfnamefont {F.}~\bibnamefont {Willaime}}, \bibinfo
  {author} {\bibfnamefont {K.}~\bibnamefont {Arakawa}}, \bibinfo {author}
  {\bibfnamefont {M.}~\bibnamefont {Gilbert}}, \ and\ \bibinfo {author}
  {\bibfnamefont {S.}~\bibnamefont {Dudarev}},\ }\href@noop {} {\bibfield
  {journal} {\bibinfo  {journal} {Physical Review B}\ }\textbf {\bibinfo
  {volume} {94}},\ \bibinfo {pages} {024103} (\bibinfo {year}
  {2016})}\BibitemShut {NoStop}%
\bibitem [{\citenamefont {Sorensen}\ \emph {et~al.}(2000)\citenamefont
  {Sorensen}, \citenamefont {Mishin},\ and\ \citenamefont
  {Voter}}]{sorensen2000}%
  \BibitemOpen
  \bibfield  {author} {\bibinfo {author} {\bibfnamefont {M.~R.}\ \bibnamefont
  {Sorensen}}, \bibinfo {author} {\bibfnamefont {Y.}~\bibnamefont {Mishin}}, \
  and\ \bibinfo {author} {\bibfnamefont {A.~F.}\ \bibnamefont {Voter}},\ }\href
  {\doibase 10.1103/PhysRevB.62.3658} {\bibfield  {journal} {\bibinfo
  {journal} {Phys. Rev. B}\ }\textbf {\bibinfo {volume} {62}},\ \bibinfo
  {pages} {3658} (\bibinfo {year} {2000})}\BibitemShut {NoStop}%
\bibitem [{\citenamefont {Uberuaga}\ \emph {et~al.}(2005)\citenamefont
  {Uberuaga}, \citenamefont {Smith}, \citenamefont {Cleave}, \citenamefont
  {Henkelman}, \citenamefont {Grimes}, \citenamefont {Voter},\ and\
  \citenamefont {Sickafus}}]{uberuaga2005}%
  \BibitemOpen
  \bibfield  {author} {\bibinfo {author} {\bibfnamefont {B.}~\bibnamefont
  {Uberuaga}}, \bibinfo {author} {\bibfnamefont {R.}~\bibnamefont {Smith}},
  \bibinfo {author} {\bibfnamefont {A.}~\bibnamefont {Cleave}}, \bibinfo
  {author} {\bibfnamefont {G.}~\bibnamefont {Henkelman}}, \bibinfo {author}
  {\bibfnamefont {R.}~\bibnamefont {Grimes}}, \bibinfo {author} {\bibfnamefont
  {A.}~\bibnamefont {Voter}}, \ and\ \bibinfo {author} {\bibfnamefont
  {K.}~\bibnamefont {Sickafus}},\ }\href@noop {} {\bibfield  {journal}
  {\bibinfo  {journal} {Nuclear Instruments and Methods in Physics Research
  Section B: Beam Interactions with Materials and Atoms}\ }\textbf {\bibinfo
  {volume} {228}},\ \bibinfo {pages} {260} (\bibinfo {year}
  {2005})}\BibitemShut {NoStop}%
\bibitem [{\citenamefont {Uberuaga}\ \emph {et~al.}(2007)\citenamefont
  {Uberuaga}, \citenamefont {Hoagland}, \citenamefont {Voter},\ and\
  \citenamefont {Valone}}]{uberuaga2007}%
  \BibitemOpen
  \bibfield  {author} {\bibinfo {author} {\bibfnamefont {B.}~\bibnamefont
  {Uberuaga}}, \bibinfo {author} {\bibfnamefont {R.}~\bibnamefont {Hoagland}},
  \bibinfo {author} {\bibfnamefont {A.}~\bibnamefont {Voter}}, \ and\ \bibinfo
  {author} {\bibfnamefont {S.}~\bibnamefont {Valone}},\ }\href@noop {}
  {\bibfield  {journal} {\bibinfo  {journal} {Physical review letters}\
  }\textbf {\bibinfo {volume} {99}},\ \bibinfo {pages} {135501} (\bibinfo
  {year} {2007})}\BibitemShut {NoStop}%
\bibitem [{\citenamefont {Perez}\ \emph {et~al.}(2009)\citenamefont {Perez},
  \citenamefont {Uberuaga}, \citenamefont {Shim}, \citenamefont {Amar},\ and\
  \citenamefont {Voter}}]{perez2009}%
  \BibitemOpen
  \bibfield  {author} {\bibinfo {author} {\bibfnamefont {D.}~\bibnamefont
  {Perez}}, \bibinfo {author} {\bibfnamefont {B.~P.}\ \bibnamefont {Uberuaga}},
  \bibinfo {author} {\bibfnamefont {Y.}~\bibnamefont {Shim}}, \bibinfo {author}
  {\bibfnamefont {J.~G.}\ \bibnamefont {Amar}}, \ and\ \bibinfo {author}
  {\bibfnamefont {A.~F.}\ \bibnamefont {Voter}},\ }\href@noop {} {\bibfield
  {journal} {\bibinfo  {journal} {Annual Reports in computational chemistry}\
  }\textbf {\bibinfo {volume} {5}},\ \bibinfo {pages} {79} (\bibinfo {year}
  {2009})}\BibitemShut {NoStop}%
\bibitem [{\citenamefont {B{\'e}land}\ \emph {et~al.}(2011)\citenamefont
  {B{\'e}land}, \citenamefont {Brommer}, \citenamefont {El-Mellouhi},
  \citenamefont {Joly},\ and\ \citenamefont {Mousseau}}]{beland2011}%
  \BibitemOpen
  \bibfield  {author} {\bibinfo {author} {\bibfnamefont {L.~K.}\ \bibnamefont
  {B{\'e}land}}, \bibinfo {author} {\bibfnamefont {P.}~\bibnamefont {Brommer}},
  \bibinfo {author} {\bibfnamefont {F.}~\bibnamefont {El-Mellouhi}}, \bibinfo
  {author} {\bibfnamefont {J.-F.}\ \bibnamefont {Joly}}, \ and\ \bibinfo
  {author} {\bibfnamefont {N.}~\bibnamefont {Mousseau}},\ }\href@noop {}
  {\bibfield  {journal} {\bibinfo  {journal} {Physical Review E}\ }\textbf
  {\bibinfo {volume} {84}},\ \bibinfo {pages} {046704} (\bibinfo {year}
  {2011})}\BibitemShut {NoStop}%
\bibitem [{\citenamefont {Voter}(1997)}]{voter1997}%
  \BibitemOpen
  \bibfield  {author} {\bibinfo {author} {\bibfnamefont {A.~F.}\ \bibnamefont
  {Voter}},\ }\href@noop {} {\bibfield  {journal} {\bibinfo  {journal}
  {Physical Review Letters}\ }\textbf {\bibinfo {volume} {78}},\ \bibinfo
  {pages} {3908} (\bibinfo {year} {1997})}\BibitemShut {NoStop}%
\bibitem [{\citenamefont {Voter}(1998)}]{voter1998parallel}%
  \BibitemOpen
  \bibfield  {author} {\bibinfo {author} {\bibfnamefont {A.~F.}\ \bibnamefont
  {Voter}},\ }\href@noop {} {\bibfield  {journal} {\bibinfo  {journal}
  {Physical Review B}\ }\textbf {\bibinfo {volume} {57}},\ \bibinfo {pages}
  {R13985} (\bibinfo {year} {1998})}\BibitemShut {NoStop}%
\bibitem [{\citenamefont {So}\ and\ \citenamefont {Voter}(2000)}]{TAD}%
  \BibitemOpen
  \bibfield  {author} {\bibinfo {author} {\bibfnamefont {M.}~\bibnamefont
  {So}}\ and\ \bibinfo {author} {\bibfnamefont {A.}~\bibnamefont {Voter}},\
  }\href@noop {} {\bibfield  {journal} {\bibinfo  {journal} {The Journal of
  Chemical Physics}\ }\textbf {\bibinfo {volume} {112}},\ \bibinfo {pages}
  {9599} (\bibinfo {year} {2000})}\BibitemShut {NoStop}%
\bibitem [{\citenamefont {Chatterjee}\ and\ \citenamefont
  {Bhattacharya}(2015)}]{chatterjee2015uncertainty}%
  \BibitemOpen
  \bibfield  {author} {\bibinfo {author} {\bibfnamefont {A.}~\bibnamefont
  {Chatterjee}}\ and\ \bibinfo {author} {\bibfnamefont {S.}~\bibnamefont
  {Bhattacharya}},\ }\href@noop {} {\bibfield  {journal} {\bibinfo  {journal}
  {The Journal of Chemical Physics}\ }\textbf {\bibinfo {volume} {143}},\
  \bibinfo {pages} {114109} (\bibinfo {year} {2015})}\BibitemShut {NoStop}%
\bibitem [{\citenamefont {Chill}\ and\ \citenamefont
  {Henkelman}(2014)}]{chill2014}%
  \BibitemOpen
  \bibfield  {author} {\bibinfo {author} {\bibfnamefont {S.~T.}\ \bibnamefont
  {Chill}}\ and\ \bibinfo {author} {\bibfnamefont {G.}~\bibnamefont
  {Henkelman}},\ }\href@noop {} {\bibfield  {journal} {\bibinfo  {journal} {The
  Journal of chemical physics}\ }\textbf {\bibinfo {volume} {140}},\ \bibinfo
  {pages} {214110} (\bibinfo {year} {2014})}\BibitemShut {NoStop}%
\bibitem [{\citenamefont {Swinburne}\ and\ \citenamefont
  {Perez}(2018{\natexlab{a}})}]{swinburne2018b}%
  \BibitemOpen
  \bibfield  {author} {\bibinfo {author} {\bibfnamefont {T.~D.}\ \bibnamefont
  {Swinburne}}\ and\ \bibinfo {author} {\bibfnamefont {D.}~\bibnamefont
  {Perez}},\ }\href {\doibase 10.1103/PhysRevMaterials.2.053802} {\bibfield
  {journal} {\bibinfo  {journal} {Phys. Rev. Materials}\ }\textbf {\bibinfo
  {volume} {2}},\ \bibinfo {pages} {053802} (\bibinfo {year}
  {2018}{\natexlab{a}})}\BibitemShut {NoStop}%
\bibitem [{\citenamefont {Swinburne}\ and\ \citenamefont
  {Perez}(2020)}]{swinburne2020c}%
  \BibitemOpen
  \bibfield  {author} {\bibinfo {author} {\bibfnamefont {T.~D.}\ \bibnamefont
  {Swinburne}}\ and\ \bibinfo {author} {\bibfnamefont {D.}~\bibnamefont
  {Perez}},\ }\href {\doibase 10.1038/s41524-020-00463-8} {\bibfield  {journal}
  {\bibinfo  {journal} {NPJ Computational Materials}\ }\textbf {\bibinfo
  {volume} {6}},\ \bibinfo {pages} {190} (\bibinfo {year} {2020})}\BibitemShut
  {NoStop}%
\bibitem [{\citenamefont {Henkelman}\ and\ \citenamefont
  {J{\'o}nsson}(1999)}]{dimer}%
  \BibitemOpen
  \bibfield  {author} {\bibinfo {author} {\bibfnamefont {G.}~\bibnamefont
  {Henkelman}}\ and\ \bibinfo {author} {\bibfnamefont {H.}~\bibnamefont
  {J{\'o}nsson}},\ }\href@noop {} {\bibfield  {journal} {\bibinfo  {journal}
  {The Journal of chemical physics}\ }\textbf {\bibinfo {volume} {111}},\
  \bibinfo {pages} {7010} (\bibinfo {year} {1999})}\BibitemShut {NoStop}%
\bibitem [{\citenamefont {Wales}(2002)}]{wales2002discrete}%
  \BibitemOpen
  \bibfield  {author} {\bibinfo {author} {\bibfnamefont {D.~J.}\ \bibnamefont
  {Wales}},\ }\href@noop {} {\bibfield  {journal} {\bibinfo  {journal}
  {Molecular physics}\ }\textbf {\bibinfo {volume} {100}},\ \bibinfo {pages}
  {3285} (\bibinfo {year} {2002})}\BibitemShut {NoStop}%
\bibitem [{\citenamefont {Leli{\`e}vre}(2018)}]{Lelievre2018}%
  \BibitemOpen
  \bibfield  {author} {\bibinfo {author} {\bibfnamefont {T.}~\bibnamefont
  {Leli{\`e}vre}},\ }\enquote {\bibinfo {title} {Mathematical foundations of
  accelerated molecular dynamics methods},}\ in\ \href {\doibase
  10.1007/978-3-319-42913-7_27-1} {\emph {\bibinfo {booktitle} {Handbook of
  Materials Modeling : Methods: Theory and Modeling}}},\ \bibinfo {editor}
  {edited by\ \bibinfo {editor} {\bibfnamefont {W.}~\bibnamefont {Andreoni}}\
  and\ \bibinfo {editor} {\bibfnamefont {S.}~\bibnamefont {Yip}}}\ (\bibinfo
  {publisher} {Springer International Publishing},\ \bibinfo {address} {Cham},\
  \bibinfo {year} {2018})\ pp.\ \bibinfo {pages} {1--32}\BibitemShut {NoStop}%
\bibitem [{\citenamefont {Henkelman}(2017)}]{henkelman2017}%
  \BibitemOpen
  \bibfield  {author} {\bibinfo {author} {\bibfnamefont {G.}~\bibnamefont
  {Henkelman}},\ }\href@noop {} {\bibfield  {journal} {\bibinfo  {journal}
  {Annual Review of Materials Research}\ } (\bibinfo {year}
  {2017})}\BibitemShut {NoStop}%
\bibitem [{\citenamefont {Le~Bris}\ \emph {et~al.}(2012)\citenamefont
  {Le~Bris}, \citenamefont {Lelievre}, \citenamefont {Luskin},\ and\
  \citenamefont {Perez}}]{lebris2012}%
  \BibitemOpen
  \bibfield  {author} {\bibinfo {author} {\bibfnamefont {C.}~\bibnamefont
  {Le~Bris}}, \bibinfo {author} {\bibfnamefont {T.}~\bibnamefont {Lelievre}},
  \bibinfo {author} {\bibfnamefont {M.}~\bibnamefont {Luskin}}, \ and\ \bibinfo
  {author} {\bibfnamefont {D.}~\bibnamefont {Perez}},\ }\href@noop {}
  {\bibfield  {journal} {\bibinfo  {journal} {Monte Carlo Methods and
  Applications}\ }\textbf {\bibinfo {volume} {18}},\ \bibinfo {pages} {119}
  (\bibinfo {year} {2012})}\BibitemShut {NoStop}%
\bibitem [{\citenamefont {Swinburne}\ and\ \citenamefont
  {Perez}(2018{\natexlab{b}})}]{tammber}%
  \BibitemOpen
  \bibfield  {author} {\bibinfo {author} {\bibfnamefont {T.}~\bibnamefont
  {Swinburne}}\ and\ \bibinfo {author} {\bibfnamefont {D.}~\bibnamefont
  {Perez}},\ }\href {https://gitlab.com/exaalt/parsplice/tree/tammber}
  {\enquote {\bibinfo {title} {\texttt{TAMMBER} branch of \texttt{ParSplice}
  code},}\ } (\bibinfo {year} {2018}{\natexlab{b}})\BibitemShut {NoStop}%
\bibitem [{\citenamefont {Swinburne}(2021)}]{swinburne2021a}%
  \BibitemOpen
  \bibfield  {author} {\bibinfo {author} {\bibfnamefont {T.~D.}\ \bibnamefont
  {Swinburne}},\ }\href@noop {} {\bibfield  {journal} {\bibinfo  {journal}
  {Computational Materials Science}\ }\textbf {\bibinfo {volume} {193}},\
  \bibinfo {pages} {110256} (\bibinfo {year} {2021})}\BibitemShut {NoStop}%
\bibitem [{\citenamefont {Opplestrup}\ \emph {et~al.}(2006)\citenamefont
  {Opplestrup}, \citenamefont {Bulatov}, \citenamefont {Gilmer}, \citenamefont
  {Kalos},\ and\ \citenamefont {Sadigh}}]{opplestrup2006first}%
  \BibitemOpen
  \bibfield  {author} {\bibinfo {author} {\bibfnamefont {T.}~\bibnamefont
  {Opplestrup}}, \bibinfo {author} {\bibfnamefont {V.~V.}\ \bibnamefont
  {Bulatov}}, \bibinfo {author} {\bibfnamefont {G.~H.}\ \bibnamefont {Gilmer}},
  \bibinfo {author} {\bibfnamefont {M.~H.}\ \bibnamefont {Kalos}}, \ and\
  \bibinfo {author} {\bibfnamefont {B.}~\bibnamefont {Sadigh}},\ }\href@noop {}
  {\bibfield  {journal} {\bibinfo  {journal} {Physical review letters}\
  }\textbf {\bibinfo {volume} {97}},\ \bibinfo {pages} {230602} (\bibinfo
  {year} {2006})}\BibitemShut {NoStop}%
\bibitem [{\citenamefont {Athenes}\ \emph {et~al.}(2019)\citenamefont
  {Athenes}, \citenamefont {Kaur}, \citenamefont {Adjanor}, \citenamefont
  {Vanacker},\ and\ \citenamefont {Jourdan}}]{athenes2019elastodiffusion}%
  \BibitemOpen
  \bibfield  {author} {\bibinfo {author} {\bibfnamefont {M.}~\bibnamefont
  {Athenes}}, \bibinfo {author} {\bibfnamefont {S.}~\bibnamefont {Kaur}},
  \bibinfo {author} {\bibfnamefont {G.}~\bibnamefont {Adjanor}}, \bibinfo
  {author} {\bibfnamefont {T.}~\bibnamefont {Vanacker}}, \ and\ \bibinfo
  {author} {\bibfnamefont {T.}~\bibnamefont {Jourdan}},\ }\href@noop {}
  {\bibfield  {journal} {\bibinfo  {journal} {Physical Review Materials}\
  }\textbf {\bibinfo {volume} {3}},\ \bibinfo {pages} {103802} (\bibinfo {year}
  {2019})}\BibitemShut {NoStop}%
\bibitem [{\citenamefont {Jourdan}\ \emph {et~al.}(2014)\citenamefont
  {Jourdan}, \citenamefont {Bencteux},\ and\ \citenamefont
  {Adjanor}}]{jourdan2014efficient}%
  \BibitemOpen
  \bibfield  {author} {\bibinfo {author} {\bibfnamefont {T.}~\bibnamefont
  {Jourdan}}, \bibinfo {author} {\bibfnamefont {G.}~\bibnamefont {Bencteux}}, \
  and\ \bibinfo {author} {\bibfnamefont {G.}~\bibnamefont {Adjanor}},\
  }\href@noop {} {\bibfield  {journal} {\bibinfo  {journal} {Journal of Nuclear
  Materials}\ }\textbf {\bibinfo {volume} {444}},\ \bibinfo {pages} {298}
  (\bibinfo {year} {2014})}\BibitemShut {NoStop}%
\bibitem [{\citenamefont {Jourdan}(2015)}]{jourdan2015influence}%
  \BibitemOpen
  \bibfield  {author} {\bibinfo {author} {\bibfnamefont {T.}~\bibnamefont
  {Jourdan}},\ }\href@noop {} {\bibfield  {journal} {\bibinfo  {journal}
  {Journal of Nuclear Materials}\ }\textbf {\bibinfo {volume} {467}},\ \bibinfo
  {pages} {286} (\bibinfo {year} {2015})}\BibitemShut {NoStop}%
\bibitem [{\citenamefont {Blondel}\ \emph {et~al.}(2017)\citenamefont
  {Blondel}, \citenamefont {Bernholdt}, \citenamefont {Hammond}, \citenamefont
  {Hu}, \citenamefont {Maroudas},\ and\ \citenamefont {Wirth}}]{Blondel2017}%
  \BibitemOpen
  \bibfield  {author} {\bibinfo {author} {\bibfnamefont {S.}~\bibnamefont
  {Blondel}}, \bibinfo {author} {\bibfnamefont {D.~E.}\ \bibnamefont
  {Bernholdt}}, \bibinfo {author} {\bibfnamefont {K.~D.}\ \bibnamefont
  {Hammond}}, \bibinfo {author} {\bibfnamefont {L.}~\bibnamefont {Hu}},
  \bibinfo {author} {\bibfnamefont {D.}~\bibnamefont {Maroudas}}, \ and\
  \bibinfo {author} {\bibfnamefont {B.~D.}\ \bibnamefont {Wirth}},\ }\href
  {\doibase 10.13182/FST16-109} {\bibfield  {journal} {\bibinfo  {journal}
  {Fusion Science and Technology}\ }\textbf {\bibinfo {volume} {71}},\ \bibinfo
  {pages} {84} (\bibinfo {year} {2017})}\BibitemShut {NoStop}%
\bibitem [{\citenamefont {Donev}\ \emph {et~al.}(2010)\citenamefont {Donev},
  \citenamefont {Bulatov}, \citenamefont {Oppelstrup}, \citenamefont {Gilmer},
  \citenamefont {Sadigh},\ and\ \citenamefont {Kalos}}]{donev2010first}%
  \BibitemOpen
  \bibfield  {author} {\bibinfo {author} {\bibfnamefont {A.}~\bibnamefont
  {Donev}}, \bibinfo {author} {\bibfnamefont {V.~V.}\ \bibnamefont {Bulatov}},
  \bibinfo {author} {\bibfnamefont {T.}~\bibnamefont {Oppelstrup}}, \bibinfo
  {author} {\bibfnamefont {G.~H.}\ \bibnamefont {Gilmer}}, \bibinfo {author}
  {\bibfnamefont {B.}~\bibnamefont {Sadigh}}, \ and\ \bibinfo {author}
  {\bibfnamefont {M.~H.}\ \bibnamefont {Kalos}},\ }\href@noop {} {\bibfield
  {journal} {\bibinfo  {journal} {Journal of Computational Physics}\ }\textbf
  {\bibinfo {volume} {229}},\ \bibinfo {pages} {3214} (\bibinfo {year}
  {2010})}\BibitemShut {NoStop}%
\bibitem [{\citenamefont {Mason}\ \emph {et~al.}(2014)\citenamefont {Mason},
  \citenamefont {Yi}, \citenamefont {Kirk},\ and\ \citenamefont
  {Dudarev}}]{mason2014}%
  \BibitemOpen
  \bibfield  {author} {\bibinfo {author} {\bibfnamefont {D.~R.}\ \bibnamefont
  {Mason}}, \bibinfo {author} {\bibfnamefont {X.}~\bibnamefont {Yi}}, \bibinfo
  {author} {\bibfnamefont {M.~A.}\ \bibnamefont {Kirk}}, \ and\ \bibinfo
  {author} {\bibfnamefont {S.~L.}\ \bibnamefont {Dudarev}},\ }\href@noop {}
  {\bibfield  {journal} {\bibinfo  {journal} {Journal of Physics: Condensed
  Matter}\ }\textbf {\bibinfo {volume} {26}},\ \bibinfo {pages} {375701}
  (\bibinfo {year} {2014})}\BibitemShut {NoStop}%
\bibitem [{\citenamefont {Jourdan}(2021)}]{jourdan2021enforcing}%
  \BibitemOpen
  \bibfield  {author} {\bibinfo {author} {\bibfnamefont {T.}~\bibnamefont
  {Jourdan}},\ }\href@noop {} {\bibfield  {journal} {\bibinfo  {journal}
  {Modelling and Simulation in Materials Science and Engineering}\ }\textbf
  {\bibinfo {volume} {29}},\ \bibinfo {pages} {035007} (\bibinfo {year}
  {2021})}\BibitemShut {NoStop}%
\bibitem [{\citenamefont {Demange}\ \emph {et~al.}(2017)\citenamefont
  {Demange}, \citenamefont {Lun{\'e}ville}, \citenamefont {Pontikis},\ and\
  \citenamefont {Simeone}}]{demange2017prediction}%
  \BibitemOpen
  \bibfield  {author} {\bibinfo {author} {\bibfnamefont {G.}~\bibnamefont
  {Demange}}, \bibinfo {author} {\bibfnamefont {L.}~\bibnamefont
  {Lun{\'e}ville}}, \bibinfo {author} {\bibfnamefont {V.}~\bibnamefont
  {Pontikis}}, \ and\ \bibinfo {author} {\bibfnamefont {D.}~\bibnamefont
  {Simeone}},\ }\href@noop {} {\bibfield  {journal} {\bibinfo  {journal}
  {Journal of Applied Physics}\ }\textbf {\bibinfo {volume} {121}},\ \bibinfo
  {pages} {125108} (\bibinfo {year} {2017})}\BibitemShut {NoStop}%
\bibitem [{\citenamefont {Noble}\ \emph {et~al.}(2020)\citenamefont {Noble},
  \citenamefont {Tonks},\ and\ \citenamefont {Fitzgerald}}]{noble2020turing}%
  \BibitemOpen
  \bibfield  {author} {\bibinfo {author} {\bibfnamefont {M.}~\bibnamefont
  {Noble}}, \bibinfo {author} {\bibfnamefont {M.}~\bibnamefont {Tonks}}, \ and\
  \bibinfo {author} {\bibfnamefont {S.}~\bibnamefont {Fitzgerald}},\
  }\href@noop {} {\bibfield  {journal} {\bibinfo  {journal} {Physical review
  letters}\ }\textbf {\bibinfo {volume} {124}},\ \bibinfo {pages} {167401}
  (\bibinfo {year} {2020})}\BibitemShut {NoStop}%
\bibitem [{\citenamefont {Li}\ \emph {et~al.}(2019)\citenamefont {Li},
  \citenamefont {Boleininger}, \citenamefont {Robertson}, \citenamefont
  {Dupuy},\ and\ \citenamefont {Dudarev}}]{li2019diffusion}%
  \BibitemOpen
  \bibfield  {author} {\bibinfo {author} {\bibfnamefont {Y.}~\bibnamefont
  {Li}}, \bibinfo {author} {\bibfnamefont {M.}~\bibnamefont {Boleininger}},
  \bibinfo {author} {\bibfnamefont {C.}~\bibnamefont {Robertson}}, \bibinfo
  {author} {\bibfnamefont {L.}~\bibnamefont {Dupuy}}, \ and\ \bibinfo {author}
  {\bibfnamefont {S.~L.}\ \bibnamefont {Dudarev}},\ }\href@noop {} {\bibfield
  {journal} {\bibinfo  {journal} {Physical Review Materials}\ }\textbf
  {\bibinfo {volume} {3}},\ \bibinfo {pages} {073805} (\bibinfo {year}
  {2019})}\BibitemShut {NoStop}%
\bibitem [{\citenamefont {Li}\ \emph {et~al.}(2020)\citenamefont {Li},
  \citenamefont {Po},\ and\ \citenamefont {Ghoniem}}]{li2020coupled}%
  \BibitemOpen
  \bibfield  {author} {\bibinfo {author} {\bibfnamefont {Y.}~\bibnamefont
  {Li}}, \bibinfo {author} {\bibfnamefont {G.}~\bibnamefont {Po}}, \ and\
  \bibinfo {author} {\bibfnamefont {N.}~\bibnamefont {Ghoniem}},\ }\href@noop
  {} {\bibfield  {journal} {\bibinfo  {journal} {Materialia}\ }\textbf
  {\bibinfo {volume} {14}},\ \bibinfo {pages} {100891} (\bibinfo {year}
  {2020})}\BibitemShut {NoStop}%
\bibitem [{\citenamefont {Yu}\ \emph {et~al.}(2021)\citenamefont {Yu},
  \citenamefont {Chatterjee}, \citenamefont {Roche}, \citenamefont {Po},\ and\
  \citenamefont {Marian}}]{yu2021coupling}%
  \BibitemOpen
  \bibfield  {author} {\bibinfo {author} {\bibfnamefont {Q.}~\bibnamefont
  {Yu}}, \bibinfo {author} {\bibfnamefont {S.}~\bibnamefont {Chatterjee}},
  \bibinfo {author} {\bibfnamefont {K.~J.}\ \bibnamefont {Roche}}, \bibinfo
  {author} {\bibfnamefont {G.}~\bibnamefont {Po}}, \ and\ \bibinfo {author}
  {\bibfnamefont {J.}~\bibnamefont {Marian}},\ }\href@noop {} {\bibfield
  {journal} {\bibinfo  {journal} {Modelling and Simulation in Materials Science
  and Engineering}\ }\textbf {\bibinfo {volume} {29}},\ \bibinfo {pages}
  {055021} (\bibinfo {year} {2021})}\BibitemShut {NoStop}%
\bibitem [{\citenamefont {McElfresh}\ \emph {et~al.}(2021)\citenamefont
  {McElfresh}, \citenamefont {Cui}, \citenamefont {Dudarev}, \citenamefont
  {Po},\ and\ \citenamefont {Marian}}]{mcelfresh2021discrete}%
  \BibitemOpen
  \bibfield  {author} {\bibinfo {author} {\bibfnamefont {C.}~\bibnamefont
  {McElfresh}}, \bibinfo {author} {\bibfnamefont {Y.}~\bibnamefont {Cui}},
  \bibinfo {author} {\bibfnamefont {S.~L.}\ \bibnamefont {Dudarev}}, \bibinfo
  {author} {\bibfnamefont {G.}~\bibnamefont {Po}}, \ and\ \bibinfo {author}
  {\bibfnamefont {J.}~\bibnamefont {Marian}},\ }\href@noop {} {\bibfield
  {journal} {\bibinfo  {journal} {International Journal of Plasticity}\
  }\textbf {\bibinfo {volume} {136}},\ \bibinfo {pages} {102848} (\bibinfo
  {year} {2021})}\BibitemShut {NoStop}%
\bibitem [{\citenamefont {Swinburne}\ and\ \citenamefont
  {Dudarev}(2018)}]{swinburne2018c}%
  \BibitemOpen
  \bibfield  {author} {\bibinfo {author} {\bibfnamefont {T.}~\bibnamefont
  {Swinburne}}\ and\ \bibinfo {author} {\bibfnamefont {S.}~\bibnamefont
  {Dudarev}},\ }\href {\doibase 10.1103/PhysRevMaterials.2.073608} {\bibfield
  {journal} {\bibinfo  {journal} {Phys. Rev. Materials}\ }\textbf {\bibinfo
  {volume} {2}},\ \bibinfo {pages} {073608} (\bibinfo {year}
  {2018})}\BibitemShut {NoStop}%
\bibitem [{\citenamefont {Kohnert}\ \emph {et~al.}(2018)\citenamefont
  {Kohnert}, \citenamefont {Wirth},\ and\ \citenamefont
  {Capolungo}}]{kohnert2018modeling}%
  \BibitemOpen
  \bibfield  {author} {\bibinfo {author} {\bibfnamefont {A.~A.}\ \bibnamefont
  {Kohnert}}, \bibinfo {author} {\bibfnamefont {B.~D.}\ \bibnamefont {Wirth}},
  \ and\ \bibinfo {author} {\bibfnamefont {L.}~\bibnamefont {Capolungo}},\
  }\href@noop {} {\bibfield  {journal} {\bibinfo  {journal} {Computational
  Materials Science}\ }\textbf {\bibinfo {volume} {149}},\ \bibinfo {pages}
  {442} (\bibinfo {year} {2018})}\BibitemShut {NoStop}%
\bibitem [{\citenamefont {Pande}\ \emph {et~al.}(2010)\citenamefont {Pande},
  \citenamefont {Beauchamp},\ and\ \citenamefont
  {Bowman}}]{pande2010everything}%
  \BibitemOpen
  \bibfield  {author} {\bibinfo {author} {\bibfnamefont {V.~S.}\ \bibnamefont
  {Pande}}, \bibinfo {author} {\bibfnamefont {K.}~\bibnamefont {Beauchamp}}, \
  and\ \bibinfo {author} {\bibfnamefont {G.~R.}\ \bibnamefont {Bowman}},\
  }\href@noop {} {\bibfield  {journal} {\bibinfo  {journal} {Methods}\ }\textbf
  {\bibinfo {volume} {52}},\ \bibinfo {pages} {99} (\bibinfo {year}
  {2010})}\BibitemShut {NoStop}%
\bibitem [{\citenamefont {Martin}(2004)}]{Martin}%
  \BibitemOpen
  \bibfield  {author} {\bibinfo {author} {\bibfnamefont {R.~M.}\ \bibnamefont
  {Martin}},\ }\href@noop {} {\emph {\bibinfo {title} {{Electronic Structure:
  Basic Theory and Practical Methods}}}}\ (\bibinfo  {publisher} {Cambridge
  University Press},\ \bibinfo {year} {2004})\BibitemShut {NoStop}%
\bibitem [{\citenamefont {Deuflhard}\ and\ \citenamefont {Weber}(2005)}]{PCCA}%
  \BibitemOpen
  \bibfield  {author} {\bibinfo {author} {\bibfnamefont {P.}~\bibnamefont
  {Deuflhard}}\ and\ \bibinfo {author} {\bibfnamefont {M.}~\bibnamefont
  {Weber}},\ }\href@noop {} {\bibfield  {journal} {\bibinfo  {journal} {Linear
  algebra and its applications}\ }\textbf {\bibinfo {volume} {398}},\ \bibinfo
  {pages} {161} (\bibinfo {year} {2005})}\BibitemShut {NoStop}%
\bibitem [{\citenamefont {R{\"o}blitz}\ and\ \citenamefont
  {Weber}(2013)}]{PCCApp}%
  \BibitemOpen
  \bibfield  {author} {\bibinfo {author} {\bibfnamefont {S.}~\bibnamefont
  {R{\"o}blitz}}\ and\ \bibinfo {author} {\bibfnamefont {M.}~\bibnamefont
  {Weber}},\ }\href@noop {} {\bibfield  {journal} {\bibinfo  {journal}
  {Advances in Data Analysis and Classification}\ }\textbf {\bibinfo {volume}
  {7}},\ \bibinfo {pages} {147} (\bibinfo {year} {2013})}\BibitemShut {NoStop}%
\bibitem [{\citenamefont {Hudson}\ \emph {et~al.}(2005)\citenamefont {Hudson},
  \citenamefont {Dudarev}, \citenamefont {Caturla},\ and\ \citenamefont
  {Sutton}}]{Hudson2005}%
  \BibitemOpen
  \bibfield  {author} {\bibinfo {author} {\bibfnamefont {T.~S.}\ \bibnamefont
  {Hudson}}, \bibinfo {author} {\bibfnamefont {S.~L.}\ \bibnamefont {Dudarev}},
  \bibinfo {author} {\bibfnamefont {M.~J.}\ \bibnamefont {Caturla}}, \ and\
  \bibinfo {author} {\bibfnamefont {A.~P.}\ \bibnamefont {Sutton}},\
  }\href@noop {} {\bibfield  {journal} {\bibinfo  {journal} {Philosophical
  Magazine}\ }\textbf {\bibinfo {volume} {85}},\ \bibinfo {pages} {661}
  (\bibinfo {year} {2005})}\BibitemShut {NoStop}%
\bibitem [{\citenamefont {Reichl}(2009)}]{Reichl2009}%
  \BibitemOpen
  \bibfield  {author} {\bibinfo {author} {\bibfnamefont {L.~E.}\ \bibnamefont
  {Reichl}},\ }\href@noop {} {\emph {\bibinfo {title} {{A Modern Course in
  Statistical Physics}}}},\ Physics Textbook\ (\bibinfo  {publisher}
  {Wiley-VCH},\ \bibinfo {year} {2009})\BibitemShut {NoStop}%
\bibitem [{\citenamefont {Ashcroft}\ and\ \citenamefont
  {Mermin}(1976)}]{Ashcroft}%
  \BibitemOpen
  \bibfield  {author} {\bibinfo {author} {\bibfnamefont {N.~W.}\ \bibnamefont
  {Ashcroft}}\ and\ \bibinfo {author} {\bibfnamefont {N.~D.}\ \bibnamefont
  {Mermin}},\ }\href@noop {} {\emph {\bibinfo {title} {{Solid state
  physics}}}},\ Holt-Saunders International Editions: Science : Physics\
  (\bibinfo  {publisher} {Holt, Rinehart and Winston},\ \bibinfo {year}
  {1976})\BibitemShut {NoStop}%
\bibitem [{\citenamefont {Swinburne}\ \emph {et~al.}(2020)\citenamefont
  {Swinburne}, \citenamefont {Kannan}, \citenamefont {Sharpe},\ and\
  \citenamefont {Wales}}]{swinburne2020d}%
  \BibitemOpen
  \bibfield  {author} {\bibinfo {author} {\bibfnamefont {T.~D.}\ \bibnamefont
  {Swinburne}}, \bibinfo {author} {\bibfnamefont {D.}~\bibnamefont {Kannan}},
  \bibinfo {author} {\bibfnamefont {D.~J.}\ \bibnamefont {Sharpe}}, \ and\
  \bibinfo {author} {\bibfnamefont {D.~J.}\ \bibnamefont {Wales}},\ }\href@noop
  {} {\bibfield  {journal} {\bibinfo  {journal} {The Journal of Chemical
  Physics}\ }\textbf {\bibinfo {volume} {153}},\ \bibinfo {pages} {134115}
  (\bibinfo {year} {2020})}\BibitemShut {NoStop}%
\bibitem [{\citenamefont {Scott}(2012)}]{scott2012group}%
  \BibitemOpen
  \bibfield  {author} {\bibinfo {author} {\bibfnamefont {W.~R.}\ \bibnamefont
  {Scott}},\ }\href@noop {} {\emph {\bibinfo {title} {Group theory}}}\
  (\bibinfo  {publisher} {Courier Corporation},\ \bibinfo {year}
  {2012})\BibitemShut {NoStop}%
\bibitem [{\citenamefont {team}(2020)}]{msmtools}%
  \BibitemOpen
  \bibfield  {author} {\bibinfo {author} {\bibfnamefont {P.}~\bibnamefont
  {team}},\ }\href {https://github.com/markovmodel/msmtools} {\enquote
  {\bibinfo {title} {\texttt{MSMTools} package},}\ } (\bibinfo {year}
  {2020})\BibitemShut {NoStop}%
\bibitem [{\citenamefont {Swinburne}\ \emph {et~al.}(2014)\citenamefont
  {Swinburne}, \citenamefont {Dudarev},\ and\ \citenamefont
  {Sutton}}]{swinburne2014}%
  \BibitemOpen
  \bibfield  {author} {\bibinfo {author} {\bibfnamefont {T.~D.}\ \bibnamefont
  {Swinburne}}, \bibinfo {author} {\bibfnamefont {S.~L.}\ \bibnamefont
  {Dudarev}}, \ and\ \bibinfo {author} {\bibfnamefont {A.~P.}\ \bibnamefont
  {Sutton}},\ }\href@noop {} {\bibfield  {journal} {\bibinfo  {journal}
  {Physical Review Letters}\ }\textbf {\bibinfo {volume} {113}},\ \bibinfo
  {pages} {215501} (\bibinfo {year} {2014})}\BibitemShut {NoStop}%
\bibitem [{\citenamefont {Swinburne}\ \emph {et~al.}(2017)\citenamefont
  {Swinburne}, \citenamefont {Ma},\ and\ \citenamefont
  {Dudarev}}]{swinburne2017}%
  \BibitemOpen
  \bibfield  {author} {\bibinfo {author} {\bibfnamefont {T.~D.}\ \bibnamefont
  {Swinburne}}, \bibinfo {author} {\bibfnamefont {P.-W.}\ \bibnamefont {Ma}}, \
  and\ \bibinfo {author} {\bibfnamefont {S.~L.}\ \bibnamefont {Dudarev}},\
  }\href@noop {} {\bibfield  {journal} {\bibinfo  {journal} {New Journal of
  Physics}\ }\textbf {\bibinfo {volume} {19}},\ \bibinfo {pages} {073024}
  (\bibinfo {year} {2017})}\BibitemShut {NoStop}%
\end{thebibliography}%

\onecolumngrid

\end{document}